# Linear dynamic polarizability and absorption spectrum of an exciton in a quantum ring in a magnetic field

**A V Ghazaryan[1], A P Djotyan[1], K Moulopoulos[2], A A Kirakosyan[1]**
[1]Department of Physics, Yerevan State University, 1 Al. Manoogian, Yerevan, Armenia
[2]Department of Physics, University of Cyprus, P.O. Box 20537, 1678 Nicosia, Cyprus

E-mail: a.ghazaryan@ysu.am, ghazaryan_areg@yahoo.com

**Abstract.** The problem of an electron-hole system interacting through a contact potential and moving in a one-dimensional quantum ring threaded by an Aharonov-Bohm flux is considered, both with respect to the system's energetics as well as its optical properties. An exact analytical expression for the energy spectrum is derived using a straightforward method based on boundary conditions for wavefunctions and their derivatives along the ring. The optical properties of this exciton system, namely the linear dynamic polarizability and absorption spectrum are investigated demonstrating certain unusual features. It is shown, for example, that for special values of the magnetic flux there are energies in the spectrum that correspond to the dark excitonic states.

## 1. Introduction

Recent success in fabrication of semiconductor nanostructures has led to the construction of III-V semiconductor volcano shaped nanorings [1-3]. Quantum correlations in such systems with interacting charged particles moving in nonsimply-connected spaces, and especially in the presence of magnetic Aharonov-Bohm (AB) fluxes, make these systems representatives of an important topic in condensed matter physics and of a new research area potentially useful for applications.

Furthermore, optical properties of confined electrons and holes in nanostructures, particularly in semiconductor quantum rings (QRs) in an external magnetic field, are a subject of increased interest in recent years [4-7]. Apart from a focus on the fundamental energetics of quasiparticles confined in QRs, investigations of optical properties of QRs have a great potential for the creation of new functional devices. In particular, the theoretical study of electric polarizability and light absorption spectrum of an exciton system is especially helpful for the development of experiments, as it provides the opportunity to compare with experimental measurement and directly verify the validity of the peculiarities of the energy levels revealed by theoretical calculations for such a multiply-connected system.

Several studies of the electron-hole system energy in a one-dimensional QR, and in the presence of a magnetic AB flux passing through the ring, were performed in the last decade [4, 7-11]. It is well known that the one-dimensional exciton system does not have a finite ground state energy for the Coulomb potential interaction [12], and in order to avoid this singularity several different models with a modified form of the interaction potential [4, 7-9] or a modified form of the dispersion law of the particles [11] in the QRs have been considered by different authors. It is known that this problem can be exactly solved only for a delta function form of the interaction potential (contact interaction) [7-9], and there are various methods to derive the exact energy spectrum condition for this particular model. In [7] this problem is solved by a series expansion method (the full wavefunction is expanded in eigenfunctions of two separate free particles, an electron and a hole), whereas in [8, 9] the same problem is solved by using a Green's function procedure.

The principal aim of this paper is to solve the problem of the exciton energy spectrum (with particles interacting through a delta function potential) in a one-dimensional QR and in the presence of the AB flux by using a more straightforward method than before; this involves a procedure of finding the eigenfunctions of the system at the points where the potential is zero, and then of imposing boundary conditions at the singular point connected with the delta function. After this we proceed to the problem of calculating the dynamic linear electric polarizability and the absorption coefficients for this system, which we believe can be very helpful for experimentalists for checking the validity of the theory developed in this paper and also in earlier investigations [7-9].

This paper is organized as follows. Section 2 contains the underlying theory and gives an exact equation that determines the entire energy spectrum of this electron-hole system. Section 3 presents the theory for the absorption spectrum of a one-dimensional exciton in a QR. Section 4 presents our results together with our main conclusions. An exactly solvable model of a finite-range potential that in a limiting case recovers our results is also given in the Appendix.

## 2. Theory

The Schrödinger equation describing the orbital physics of a system of an electron and a hole (assumed spinless) interacting by a delta function potential in the presence of the AB flux in a one-dimensional quantum nanoring can be written in the form

$$\left[\frac{1}{2m_1}\left(-i\hbar\frac{\partial}{R\partial\varphi_1}-\frac{q_1}{c}A\right)^2+\frac{1}{2m_2}\left(-i\hbar\frac{\partial}{R\partial\varphi_2}-\frac{q_2}{c}A\right)^2+U\delta(\varphi_1-\varphi_2)\right]\psi(\varphi_1,\varphi_2)=E\psi(\varphi_1,\varphi_2), \qquad (2.1)$$

where $m_1(m_2)$ and $q_1(q_2)$ are the first (second) particle mass and charge respectively, $R$ is the ring radius, $U$ is the interaction potential constant and $\mathbf{A}$ is the vector potential, which is chosen to be tangential to every point of the circular ring with a constant magnitude $A=\Phi/2\pi R$. (This is the simplest gauge choice for an AB situation (where the magnetic flux $\Phi$ does not hit the one-dimensional path), but it is also valid for a case of a magnetic field $B$ that actually reaches the ring's circumference and the particles themselves, provided that the field $B$ is homogeneous (and then $\Phi=\pi BR^2$)). In order to solve (2.1), we first bring it into a decoupled form by use of the center of mass $(\Phi_c)$ and relative $(\varphi)$ variables, defined by

$$\varphi=\varphi_1-\varphi_2, \qquad \Phi_c=\frac{m_1\varphi_1+m_2\varphi_2}{m_1+m_2}. \qquad (2.2)$$

After writing the wave function of the system $\psi(\varphi_1,\varphi_2)$ in the form $\psi(\varphi_1,\varphi_2)=\Psi_c(\Phi_c)\Phi(\varphi)$, the decoupled Schrödinger equation is transformed to

$$\left[\frac{1}{2M_c}\left(-i\hbar\frac{\partial}{R\partial\Phi_c}-\frac{Q_c}{c}A\right)^2+\frac{1}{2\mu_r}\left(-i\hbar\frac{\partial}{R\partial\varphi}-\frac{Q_r}{c}A\right)^2+U\delta(\varphi)\right]\Psi_c(\Phi_c)\Phi(\varphi)=E\Psi_c(\Phi_c)\Phi(\varphi), \quad (2.3)$$

where

$$M_c=m_1+m_2,\ \mu_r=\frac{m_1 m_2}{m_1+m_2},\ Q_c=q_1+q_2,\ Q_r=\frac{q_1 m_2-q_2 m_1}{m_1+m_2}. \quad (2.4)$$

The total energy of the system $E$ can be written as $E=E_r+E_c$, where $E_r$ is the energy of relative motion and $E_c$ is the center of mass energy. The center of mass represents a free particle (with mass $M_c$ and charge $Q_c$) and this problem has the following solutions:

$$E_c=\frac{\hbar^2}{2M_c R^2}\left(N-\frac{Q_c}{hc}\Phi\right)^2, \quad \Psi_c(\Phi_c)=\frac{1}{\sqrt{2\pi}}\exp(iN\Phi_c),\ N=0,\pm 1,\pm 2,... \quad (2.5)$$

(with $N$ being related to the eigenvalues of the orbital angular momentum $L_z$ of the whole pair). The problem of the relative motion, after using the definitions

$$B=\frac{E_r}{\Delta}=\frac{E-E_c}{\Delta},\quad f=\frac{Q_r\Phi}{hc},\quad \Delta=\frac{\hbar^2}{2\mu_r R^2}, \quad (2.6)$$

can be written in the form

$$\left[\left(-i\frac{\partial}{\partial\varphi}-f\right)^2-B\right]\Phi(\varphi)=-\frac{U}{\Delta}\delta(\varphi)\Phi(\varphi). \quad (2.7)$$

It should be noted here that in (2.7) the potential is infinite only for $\varphi=0$, which is only true if $\varphi$ is defined to vary in the interval $-\pi\leq\varphi<\pi$. The potential on the ring, however, should actually be $2\pi$-periodic if we allow for $-\infty\leq\varphi<\infty$, and in this interval the potential should be in the form $V(\varphi)=U\sum_{n=-\infty}^{\infty}\delta(\varphi-2\pi n)$. After using this full expression of the potential in (2.7) and after making a gauge transformation of the form

$$\Phi(\varphi)=\exp(if\varphi)\chi(\varphi), \quad (2.8)$$

the Schrödinger equation for the relative motion becomes

$$\left(\frac{\partial^2}{\partial\varphi^2}+B\right)\chi(\varphi)=\frac{U}{\Delta}\sum_{n=-\infty}^{\infty}\delta(\varphi-2\pi n)\,\chi(\varphi). \quad (2.9)$$

Because of the periodicity of the potential in (2.9), the function $\chi(\varphi)$ must have the Bloch form, namely

$$\chi(\varphi)=\exp(ik\varphi)u(\varphi), \quad (2.10)$$

where $u(\varphi)$ is a periodic function of $\varphi$ with the same period $2\pi$, and $k$ is a real dimensionless number (the analog of "crystal momentum") that lies within the first Brillouin zone, namely

$$-\frac{1}{2}<k\leq\frac{1}{2}. \quad (2.11)$$

By using (2.5), (2.8) and (2.10) the full wavefunction can be written in terms of the separate variables of each particle in the form

$$\psi(\varphi_1,\varphi_2)=\psi(\varphi,\Phi_c)=C\exp\left(iN\frac{m_1\varphi_1+m_2\varphi_2}{m_1+m_2}\right)\exp\left(i(f+k)(\varphi_1-\varphi_2)\right)u(\varphi_1-\varphi_2). \tag{2.12}$$

By then imposing the usual singlevaluedness on the ring, namely that the full wavefunction should be $2\pi$ - periodic for each variable separately, we obtain additional conditions on the parameters contained in (2.12), namely

$$\frac{m_1 N}{m_1+m_2}+f+k=n_1, \qquad \frac{m_2 N}{m_1+m_2}-f-k=n_2, \tag{2.13}$$

where $n_1$ and $n_2$ are arbitrary integers. From (2.13) it follows that $N=n_1+n_2$, namely the constant $N$ ($\hbar N$ is the eigenvalue of the total angular momentum in $z$ direction) is an integer, and $f+k=\frac{m_2}{m_1+m_2}n_1-\frac{m_1}{m_1+m_2}n_2=l$. From all the above we conclude that the relative wavefunction has the form

$$\Phi(\varphi)=\exp(il\varphi)u(\varphi) \tag{2.14}$$

with $u(\varphi)$ being $2\pi$ - periodic. This property will be used in the following in order to obtain the proper boundary conditions that will have to be imposed on the wavefunction for solving the Schrödinger equation.

In order to derive those boundary conditions let us find the relative wave function in the region $0<\varphi<2\pi$, where the potential is zero. In this region (2.7) becomes the simple equation for a free particle (of mass $\mu_r$ and charge $Q_r$), with the solution having the form

$$\Phi(\varphi)=e^{if\varphi}\left(C_1 e^{i\sqrt{B}\varphi}+C_2 e^{-i\sqrt{B}\varphi}\right), \tag{2.15}$$

where $C_1$ and $C_2$ are the constants which will be derived from boundary conditions and from normalization condition. By using (2.14) and (2.15) it is easy to derive the expression for the wave function in the region $2\pi<\varphi<4\pi$. In order to get the wavefunction for every value of the relative variable $\varphi$ we should impose the boundary conditions at the point $\varphi=2\pi$, where the potential is present, by matching from the two regions where the wavefunction is nonvanishing. The first boundary condition is the usual wave function continuity condition, namely:

$$\Phi(\varphi)\big|_{\varphi=2\pi^-}=\Phi(\varphi)\big|_{\varphi=2\pi^+}, \tag{2.16}$$

which, with the use of the (2.14), can be written in the form

$$\Phi(\varphi)\big|_{\varphi=2\pi^-}=e^{i2\pi l}\Phi(\varphi)\big|_{\varphi=0^+}. \tag{2.17}$$

Because of the delta form of the potential, the derivative of the wavefunction is discontinuous. In order to find the condition that the derivative of the wavefunction should satisfy, (2.7) should be integrated in the region $\{-\eta,\eta\}$, and then by taking the limit $\eta\to 0$ we arrive at the relation

$$\Phi'(\varphi)\big|_{\varphi=0^+}=\Phi'(\varphi)\big|_{\varphi=0^-}+\frac{U}{\Delta}\Phi(0), \tag{2.18}$$

which, again with the use of (2.14), can be written in the form

$$\Phi'(\varphi)\big|_{\varphi=0^+}=e^{-i2\pi l}\Phi'(\varphi)\big|_{\varphi=2\pi^-}+\frac{U}{\Delta}\Phi(0). \tag{2.19}$$

Finally, by using (2.15) in (2.17) and (2.19), we obtain a system of two homogenous linear equations, that will have a solution if the corresponding determinant is zero, which after certain analytical manipulations gives

$$\cos(2\pi(f-l)) - \cos(2\pi\sqrt{B}) = \frac{U}{2\Delta\sqrt{B}}\sin(2\pi\sqrt{B}). \qquad (2.20)$$

This is the energy defining equation for the relative (internal pair-) system and we should note that it is the same condition as the one obtained for the problem of the exciton spectrum in a QR in the same AB setting worked out in other investigations [7, 8] using totally different methods. For the corresponding relative eigenfunctions we get the expression

$$\Phi(\varphi) = Ce^{if\varphi}\left(\cos(\sqrt{B}\varphi) - e^{2i\pi(f-l)}\cos[\sqrt{B}(\varphi-2\pi)] + \frac{U}{\Delta\sqrt{B}}\sin[\sqrt{B}\varphi]\right), \qquad (2.21)$$

and from the normalization condition we obtain that the normalization constant is

$$C = (2\sqrt{B})^{1/2}\left\{4\pi\sqrt{B} - \cos[2\pi(f-l)]\left(4\pi\sqrt{B}\cos(2\pi\sqrt{B}) + 2\sin(2\pi\sqrt{B})\right) + \sin(4\pi\sqrt{B})\right.$$

$$\left. + \frac{U}{\Delta\sqrt{B}} + \frac{2\pi U^2}{\Delta^2\sqrt{B}} - \frac{U}{\Delta\sqrt{B}}\cos(4\pi\sqrt{B}) - \frac{4\pi U}{\Delta}\cos[2\pi(f-l)]\sin(2\pi\sqrt{B}) - \frac{U^2}{2\Delta^2 B}\sin(4\pi\sqrt{B})\right\}^{-1/2}.$$

(2.22)

## 3. The absorption spectrum of an exciton in a QR

After finding the energy defining equation for the two-particle system, let us now switch our attention to the response of the system to an external electric field, i.e. by addressing the issue of the dynamic linear polarizability and the absorption spectrum. In calculations of optical properties, it is commonly assumed that the ring has finite thickness and a square cross-section with linear size $L < a_B^*$ ($a_B^* = \hbar^2\varepsilon/m_e e^2$ is the effective Bohr radius in bulk semiconductor, with $\varepsilon$ being the static dielectric constant of the bulk material). As in the case of the quantum wire, the choice of the cross section shape does not have any essential influence in the case of equal cross sections areas [13]. Due to the small thickness in comparison to the radius of the ring and to the characteristic length of bound states, we assume that the interaction between particles is one-dimensional and takes place only along the perimeter of the ring (also, we assume that the confining potential creates appropriate infinite barriers for the electron and the hole on the ring surface).

We then consider the absorption of a monochromatic electromagnetic wave with the frequency close to the transition of the electron from the valence band to the ground and excited excitonic states connected with the first subband created by the ring lateral confinement.

As an initial state for the system we take the state where the electron and the hole are recombined with each other and we do not have any free particle. The expression for the dynamic linear polarizability then has the form [14]

$$\chi(\omega) = -|d_{cv}|^2 \sum_\mu |\Phi_\mu(\mathbf{r}=0)|^2 \left(\frac{1}{\hbar\omega - E_g^* - E_{r,\mu} + i\delta} - \frac{1}{\hbar\omega + E_g^* + E_{r,\mu} + i\delta}\right), \qquad (3.1)$$

where $\mathbf{r} = \mathbf{r}_1 - \mathbf{r}_2$ (with $\mathbf{r}_1$, $\mathbf{r}_2$ being the radius vectors of the first and second particle respectively), $\mu$ is the quantum number describing the quantum states of the relative motion and $d_{cv}$ is the electric dipole matrix element between bands defined by

$$d_{cv} = \int d\mathbf{r}' \psi_c^*(\mathbf{r}') Q_r R \varphi' \psi_v(\mathbf{r}'), \qquad (3.2)$$

$\delta$ is the switch-on parameter described in [14] and $E_g^*$ is the effective band gap energy, which includes the energy of the lateral geometrical confinement of the electron and the hole.

In (3.1) we have the summation over all energies of the exciton defined by the solutions of (2.20), and the two terms in the brackets describe the resonant and the non-resonant terms of the absorption. It should be noted here that for particles with nonzero value of $Q_c$, there is an additional contribution to the linear polarizability from center of mass motion, which we will ignore; in the graphics that are presented in the next section we take $Q_c = 0$. It should be stressed that in (3.1) there should be an additional integral because of the center of mass wavefunction, which after integration gives the condition $\delta_{N,0}$, this meaning that in the summation of (3.1) only states with $N = 0$ actually contribute.

Using the commutation relations between the variable $\varphi$ and the Hamiltonian the dipole matrix element can be written in the form

$$|d_{cv}|^2 = \frac{\hbar^2 Q_r^2}{m_0^2 E_g^2} |P_{cv}|^2, \qquad (3.3)$$

where $P_{cv}$ is a one electron $p$-matrix element between valence and conduction bands, and $m_0$ is the bare electron mass.

Although the linear polarizability of (3.1) contributes to the refractive index $n(\omega)$ of the system as a whole, its contribution is not substantial and we can make the assumption that the refractive index of the system as a whole is not dependent on the electric field frequency $\omega$ for which the exciton absorption is observed. So, we can take $n(\omega) = n_b$, where $n_b = \sqrt{\varepsilon}$ is the refractive index.

Finally, it is well known that the absorption coefficient is proportional to the imaginary part of the polarizability; using (3.1) and (3.3) the absorption coefficient and the real part of the dynamic linear polarizability can therefore be written in the compact forms

$$\alpha(\omega) = \frac{4\pi\omega}{\sqrt{\varepsilon} c} \operatorname{Im} \chi(\omega) = \frac{4\pi^2 \hbar^2 \omega Q_r^2}{\sqrt{\varepsilon} c} \frac{|P_{cv}|^2}{m_0^2 E_g^2} \sum_\mu |\Phi_\mu(\varphi=0)|^2 \delta(\hbar\omega - E_g^* - E_{r,\mu}), \qquad (3.4)$$

$$\operatorname{Re} \chi(\omega) = \frac{2 Q_r^2 |P_{cv}|^2 \hbar^2}{m_0^2 E_g^2} \sum_\mu |\Phi_\mu(\varphi=0)|^2 \frac{E_g^* + E_{r,\mu}}{\left(E_g^* + E_{r,\mu}\right)^2 - (\hbar\omega)^2}. \qquad (3.5)$$

Equations (2.20) - (2.22) lead to the special dependence of energies and wavefunctions on the AB flux, which if substituted in (3.4) and (3.5) can directly yield the required optical properties. This, together with graphical representations of all quantities, is carried out in the following section.

## 4. Results and discussion

We apply the above results to a quantum ring made of GaAs (and for the calculation of optical properties we assume a square cross section of the ring with linear size $L$). Accordingly, the parameters used in this paper have the following values: $m_e = 0.067 m_0$, $m_{lh} = 0.087 m_0$, $E_g = 1.424 eV$,

$\varepsilon = 13.18$ [15], $P_L^2 = |P_{cv}|^2/m_0 = 14.4 eV$ [16]. With these parameters the effective Bohr radius is $a_B^* = \hbar^2\varepsilon/m_e e^2 = 103.44\text{Å}$ and the effective Rydberg energy is $E_R^* = m_e e^4/2\hbar^2\varepsilon^2 = 5.28 meV$. For the radius and thickness of the ring we take values that are comparable to those observed in experiment [17], namely $R = 2a_B^*$, $L = 2a_B^*/5$.

In figure 1 the dependence of the energy levels on magnetic flux for ground as well as for excited states for different $N$ is shown. Here we take that $q_1 = -q_2 = e$, so that $Q_c = 0$, $Q_r = e$, and $f = \Phi/\Phi_0$, where $\Phi_0 = hc/e$ is the magnetic flux quantum. In figure 1 the dimensionless energy is defined in units of $\Delta$, which in terms of $E_R^*$ has the value

$$\Delta = \frac{1.77}{R_0^2} E_R^*, \quad \text{where } R_0 = \frac{R}{a_B^*}; \tag{4.1}$$

for a ring radius equal to $R_0 = 2$, $\Delta = 0.443 E_R^* = 2.34 meV$, $L \approx 41.38\text{Å}$.

As can be seen from figure 1 the ground and excited states show the expected AB oscillations with variation of $f$ from 0 to 1, but only for the $N = 0$ case these oscillations are symmetric around the point $f = 0.5$. This occurs due to the fact that electron and hole have different masses and in (2.20) $l$ is not an integer. Figure 1a shows that even for $N = 1$ there are some values of $f$ for which the energy is negative. It should be noted, however, that there is no need for the total energy to be negative in order to form a bound state; it is sufficient to have $E_r < 0$, from which it turns out that bound states appear whenever $U < U_c$, with the critical value of interaction being

$$U_c = -\frac{\Delta}{\pi}\left(1 - \cos\left[2\pi\left(f + \frac{m_{lh}}{m_e + m_{lh}}N\right)\right]\right). \tag{4.2}$$

The condition $U < U_c$ is always satisfied for the $U = -\Delta$, so that all three ground states shown in figure 1a correspond to bound states. Such bound states are also shown in figure 2a for sufficiently attractive interaction strength $U$, while for a smaller attraction (for $U = -0.5\Delta$) a transition from bound to unbound states is observed for a range of values of the flux parameter $f$ (for which it so happens that $U \geq U_c$). Indeed, for $N = 0$ the energy connected with the center of mass is zero, the total energy of the system is therefore equal to the energy of relative motion and, as can be seen from figure 2a, for $U = -0.5\Delta$ and for this window of $f$ - values the energy becomes positive, signifying a transition to unbound states in that window. Figure 2b reveals, however, a more interesting feature of this system: for values of $f = 0.5$ for the first (and in the case $f = 0$ for the second) excited state, the energy levels do not depend on the value of the interaction potential and they have constant values. These states are formed when the difference $\cos(2\pi(f-l)) - \cos(2\pi\sqrt{B})$ and $\sin(2\pi\sqrt{B})$ in (2.20) become zero simultaneously, so that (2.20) is satisfied regardless of the value of the interaction potential. This condition is satisfied when $f - l = n/2$, where $n$ is an arbitrary integer, and for these values of $f$ the solution of (2.20) turns out to be $B = \left(m + \dfrac{n}{2}\right)^2$, where $m$ is another integer. These are essentially free particle energies (for the relative system) and their existence originates from the particular form of the wavefunctions whenever the above condition holds: it is easy to see from (2.21) that for these energy values the wavefunction becomes zero at $\varphi = 0$. This means that, for these special energies, the "relative particle" that corresponds to (2.7) does not have any contact with the potential and is therefore completely free. This property will have its impact on linear dynamic polarizability and absorption spectrum as will be shown below, because these quantities will be shown to be directly connected with $\Phi(\varphi = 0)$, which is zero for this case.

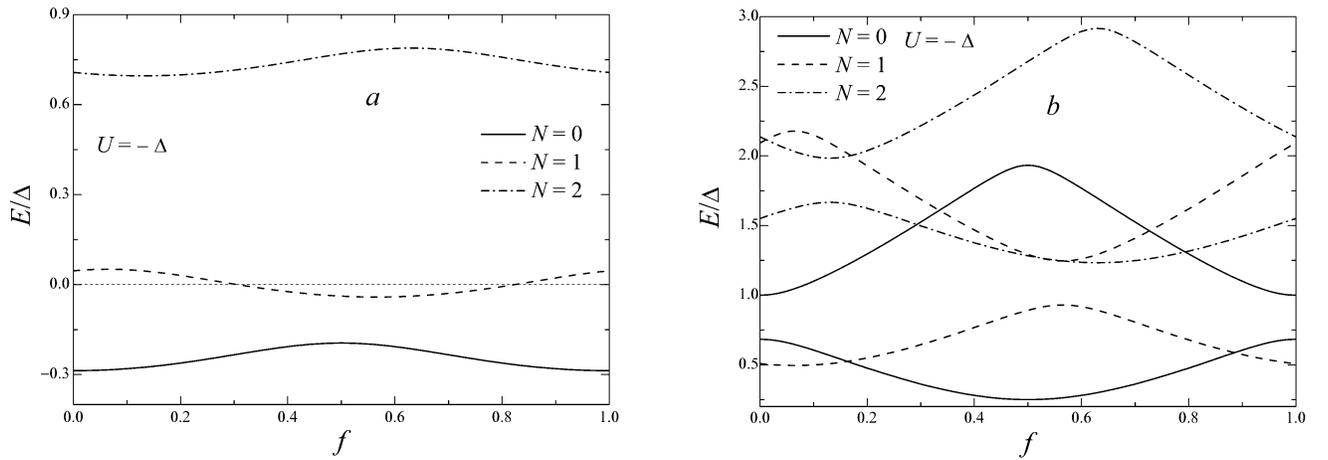

**Figure 1.** The dependence of the ground (a) and excited (b) states energies on magnetic flux parameter $f$ for different values of $N$ of total angular momentum.

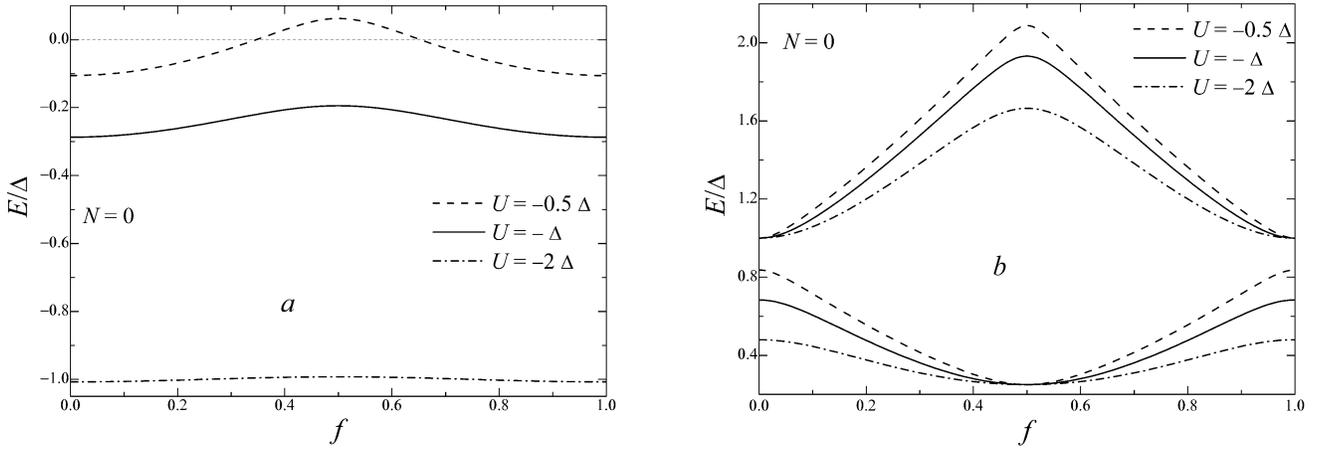

**Figure 2.** The dependence of the ground (a) and excited (b) states energies on flux parameter $f$ for different values of parameter $U$ and for $N = 0$.

From the above discussion it is expected that the above feature will not be observed for finite range potentials (the delta-function form was essential), but for short range potentials it may still be possible to observe a similar behaviour. The relative wavefunction for unbound states is an oscillating function, and if the width of the potential is much smaller than the period of oscillation of the wavefunction (which is about $2\pi/\sqrt{B}$), this effect will to some extent be observable in such systems as well.

Since the AB flux adds to the phase of the wavefunction, it is clear that, by varying the flux, the phase of the oscillations of the wavefunction will change and the particle will now be able to interact with the delta potential and the above effect will disappear, in accordance to what is observed in figure 2b.

Let us now turn on the consequences of the above to the optical properties. As can be seen from (3.4), for states with vanishing wavefunction at $\varphi = 0$, the absorption coefficient is equal to zero (this can also be seen in figure 4 below). This means that, for this idealized model, the absorption (transition) from an initial state to these special states is forbidden. In realistic systems (but similar to our model problem), once these exciton states are created, they will have long lifetime compared to all other states, and they will be technically dark in the emission spectra; by then following [5] it is reasonable to call these states "dark excitonic states", as opposed to the other ones that can be called "bright".

The dependence of the linear dynamic polarizability on photon energy is shown in figure 3 for different values of flux parameter $f$. As already discussed, the "relative particle" probability density for $\varphi = 0$ becomes zero for specific values of $f$, and as can be seen from figure 3 this has a direct impact on the linear polarizability (see (3.5)) especially on the number of singular points of $\text{Re}\,\chi(\omega)$.

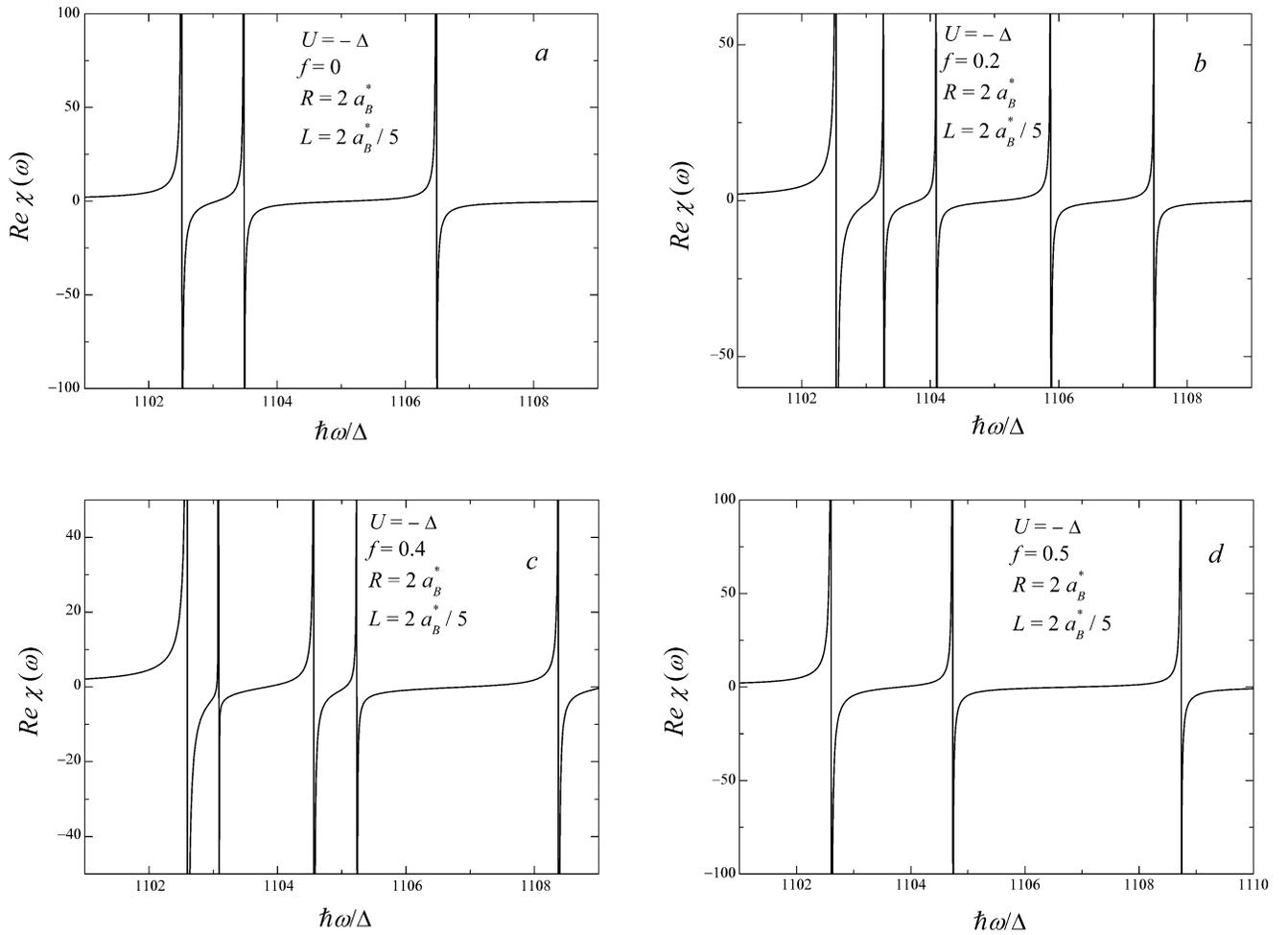

**Figure 3.** The dependence of the linear dynamic polarizability of an exciton in a QR (with $R = 2a_B^*$, $L = 2a_B^*/5$) on the photon energy for different values of the flux parameter $f$.

Figure 4 shows the values of the absorption coefficient for several states and for different values of the flux parameter $f$. As can be seen from figure 4 the absorption coefficient of the first bound state is the highest, as expected, because only this state corresponds to the bound state of the exciton.

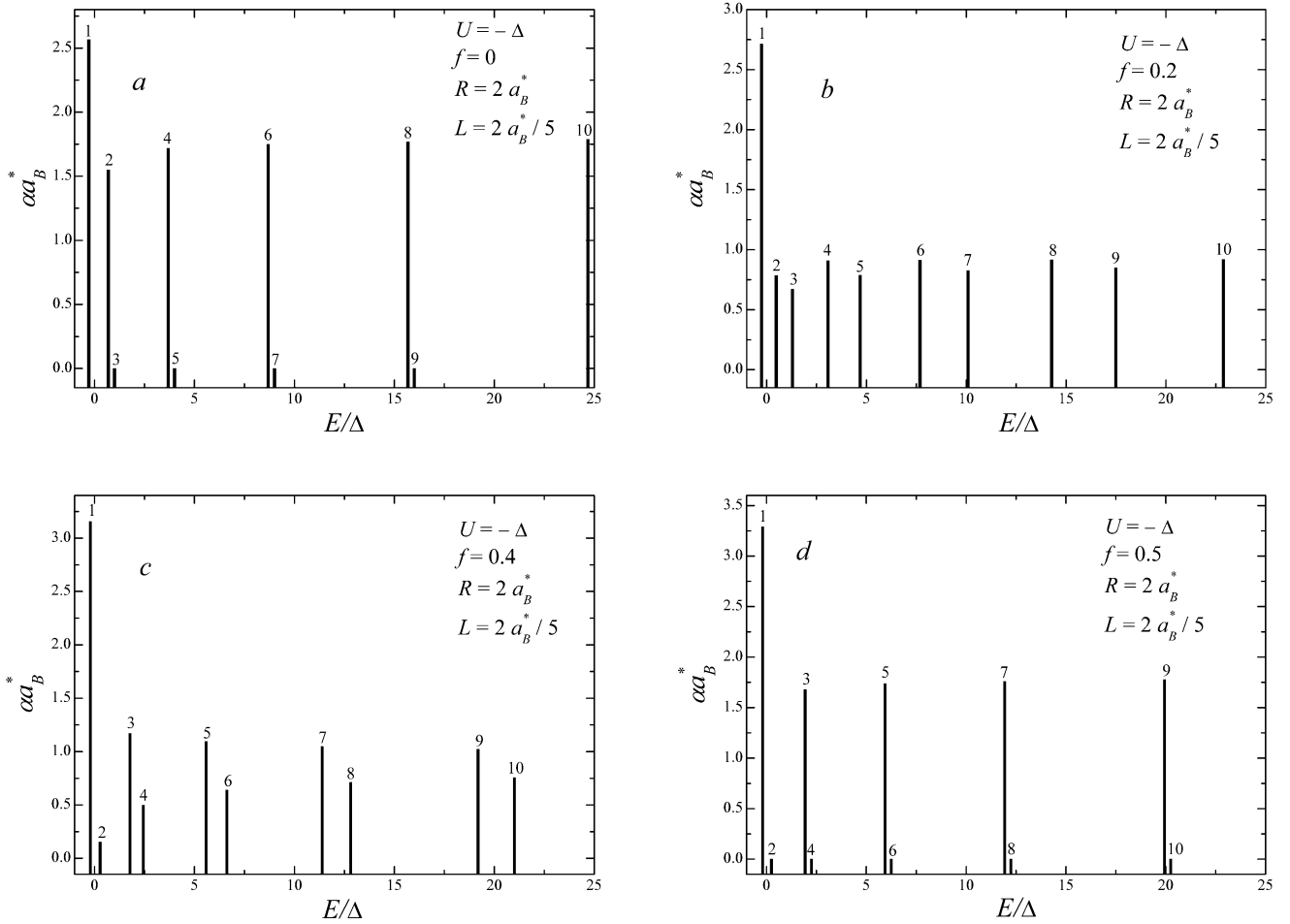

**Figure 4.** The values of the excitonic absorption coefficient in the QR with $R = 2a_B^*$, $L = 2a_B^*/5$ for several states and for different values of the flux parameter $f$.

For $f = 0$, the absorption spectra of the exciton excited states in the QR show a dependence on energy that is similar to the one observed for the exciton continuum states in the bulk material. Although the excited states in the ring are discrete due to the QR geometry, these states are nevertheless unbound and it is quite natural that the behaviour of the optical characteristics of the QR resembles the one of the bulk system.

From figure 4a it is clearly seen that in the $f = 0$ case, four states (3,5,7,9) have zero absorption coefficient and, as predicted above, this is the direct consequence of the delta potential interaction. Furthermore, it is observed that the presence of the magnetic flux changes the values of absorption coefficients drastically. As shown in figure 4 (a and b), the ratio between the absorption coefficient values for transitions to the ground and first excited excitonic states for $f = 0.2$ becomes larger compared to that for the $f = 0$ case. This comes from the fact that the magnetic flux has the tendency to move the electron and the hole in different directions and, consequentially, the probability of finding the electron and the hole at the same point for unbound states will become smaller. Furthermore, the variation of magnetic flux from $f = 0$ to $f = 0.5$ changes the absorption coefficient for some states from zero to the maximal value, while for other states this occurs in the opposite manner. As already stated, the wavefunction shows oscillating behavior, and the adjacent excited states (with assigned numbers $n$ and $n+1$ in figure 4) differ from each other essentially by a phase $\pi/2$. This is the reason why, when the wavefunction is zero at $\varphi = 0$, we have the appearance of the maximum of the adjacent states wavefunctions, as seen in figure 4a. And as already mentioned, the wavefunction acquires an additional phase in the magnetic field, and because of this, variation of $f$ by 0.5 transforms the dark excitonic states to bright and correspondingly bright excitonic states to dark ones.

## 5. Conclusions

In this work we considered the problem of an electron and a hole interacting through contact potential in a quantum ring (with square cross section and infinite potential barrier on the ring surface) threaded by a magnetic Aharonov-Bohm flux. It was shown that this problem can be divided into two separate problems: the center of mass and the relative motion. The center of mass problem corresponds to a free particle and has trivial solution. For solving the relative problem we have used an alternative method (compared to the ones used earlier by other authors [7-9]) that is more straightforward and easier to follow. Using the wave function and energy levels found before, we calculated the optical characteristics for this excitonic system, namely dynamic linear polarizability and absorption spectrum. It was shown that for special values of magnetic flux and for several excited states, the energy does not depend on the value of interaction potential. This phenomenon was explained from the fact that at these points the relative particle does not feel the influence of the interaction potential and its behaviour corresponds to a free particle. It was shown that the optical spectrum of these states corresponds to the dark excitonic states and that they have a direct impact on the dynamic polarizability. This was a consequence of the nature of the interaction potential (a delta function), but it was argued that this effect can also be partially observed in the case of a sufficiently short range potential.

Finally, the analogue of the above problem for a finite range square well potential interaction can also be solved exactly, and as shown in the Appendix, the above results for a contact interaction are recovered by a proper limiting procedure.

**Appendix**

Let us briefly discuss the problem of an exciton in a one dimensional quantum ring with the interaction potential approximated by a finite square well. We will show that the above obtained results with delta potential are an appropriate limiting case of this problem.

For the potential we take the form

$$V(\varphi) = \begin{cases} -V & -\varphi_0 + 2\pi k \leq \varphi \leq \varphi_0 + 2\pi k \\ 0 & \varphi_0 + 2\pi k < \varphi < 2\pi - \varphi_0 + 2\pi k \end{cases} \quad k = 0, \pm 1, \pm 2... \quad , \quad (A.1)$$

where $\varphi_0 = a/2R$ and $a$ is the width of the well. In order to solve this problem let us take the region $-\varphi_0 \leq \varphi \leq 2\pi - \varphi_0$ and divide it into two different parts, where the solution can be written explicitly. For the first part ($-\varphi_0 \leq \varphi \leq \varphi_0$) the problem of an exciton becomes the problem of a particle in the field of the constant potential $-V$, for which the solution has the form

$$\Phi_1(\varphi) = C_1 e^{i\left(\sqrt{B+V_0}+f\right)\varphi} + C_2 e^{-i\left(\sqrt{B+V_0}-f\right)\varphi}, \quad (A.2)$$

where $V_0 = V/\Delta$, and we have used notations as above. For the second region ($\varphi_0 \leq \varphi \leq 2\pi - \varphi_0$) the problem is the same as that of a free particle and the solution has the form

$$\Phi_2(\varphi) = C_1 e^{i\left(\sqrt{B}+f\right)\varphi} + C_2 e^{-i\left(\sqrt{B}-f\right)\varphi}. \quad (A.3)$$

By using the two forms of the wavefunction and then by imposing boundary conditions (which for this case are the continuity conditions for wave functions and also for the derivatives) at the points $\varphi = \varphi_0$ and $\varphi = 2\pi - \varphi_0$, we get the system of four homogenous linear equations, which will have a nontrivial solution whenever the determinant of these equations is zero. After some tedious calculations this gives the energy defining equation that turns out to be

$$\begin{aligned}2\sqrt{B}\sqrt{B+V_0}\left(\cos\left[2\pi(f-l)\right] - \cos\left[2\sqrt{B+V_0}\varphi_0\right]\cos\left[2\sqrt{B}(\pi-\varphi_0)\right]\right) = \\ -(2B+V_0)\sin\left[2\sqrt{B+V_0}\varphi_0\right]\sin\left[2\sqrt{B}(\pi-\varphi_0)\right]\end{aligned}. \quad (A.4)$$

In order to get the limiting case for the delta potential we must take that $V_0 \to \infty$, $\varphi_0 \to 0$ with the condition that the quantity $Va/R = -U$ is constant, and this transforms (A.4) into (2.20).

**References**

[1]   Garsia J M, Medeiros-Ribeiro G, Schmidt K, Ngo T, Feng J L, Lorke A, Kotthaus J and Petroff P M 1997 *Appl. Phys. Lett.* **71** 2014
[2]   Raz T, Ritter D and Bahir G 2003 *Appl. Phys. Lett.* **82** 1706
[3]   Climente J I and Planelles J 2008 *Journ. of Phys.: Condens. Matter* **20** 035212
[4]   Hu H, Zhu J L, Li D J and Xiong J J 2001 *Phys. Rev.* B **63** 195307
[5]   Govorov A O, Ulloa S E, Karrai K and Warburton R J 2002 *Phys. Rev.* B **66** 081309
[6]   Song J and Ulloa S E 2001 *Phys. Rev.* B **63** 125302
[7]   Römer R A and Raikh M E 2000 *Phys. Rev.* B **62** 7045
[8]   Moulopoulos K and Constantinou M 2004 *Phys. Rev.* B **70** 235327
      Moulopoulos K and Constantinou M 2007 *Phys. Rev.* B **76** 039902 (erratum)
[9]   Kyriakou K, Moulopoulos K, Ghazaryan A V and Djotyan A P 2010 *J. Phys. A: Math. Theor.* **43** 354018 (Preprint cond-mat/1007.0768)
[10]  Chaplik A V 1995 *JETP Lett.* **62** 900
[11]  Avetisyan A A, Ghazaryan A V, Djotyan A P, Kirakosyan A A and Moulopoulos K 2009 *Acta Physica Polonica* A **116** 826
[12]  Loudon R 1959 *Am. J. Phys.* **27** 649
[13]  Bryant G 1985 *Phys. Rev.* B **31** 7812
[14]  Haug H and Koch S W 2004 *Quantum Theory of the Optical and Electronic Properties of Semiconductors* (Singapore: World Scientific Publishing)
[15]  Adachi S 1985 *J. Appl. Phys.* **58** R1
[16]  Vurgaftman I, Meyer J R and Ram-Mohan L R 2001 *J. Appl. Phys.* **89** 5815
[17]  Ding F et al 2010 *Phys. Rev.* B **82** 075309